\documentclass[usenatbib]{mnras}

\usepackage{newtxtext,newtxmath}
\usepackage{float}

\usepackage[T1]{fontenc}

\DeclareRobustCommand{\VAN}[3]{#2}
\let\VANthebibliography\thebibliography
\def\thebibliography{\DeclareRobustCommand{\VAN}[3]{##3}\VANthebibliography}

\usepackage{graphicx}	
\usepackage{amsmath}	
\usepackage{amssymb}	

\title[Density-Ionization Correlations in the 21cm Power Spectrum]{Quantifying Density-Ionization Correlations with the 21cm Power Spectrum}

\author[Pagano \& Liu]{
Michael Pagano,$^{1}$\thanks{E-mail: michael.pagano@mail.mcgill.ca}
Adrian Liu$^{1}$
\\
$^{1}$Department of Physics and McGill Space Institute, McGill University, Montreal, QC, Canada H3A 2T8\\
}

\date{Submitted May 21st, 2020}

\begin{document}
\label{firstpage}
\pagerange{\pageref{firstpage}--\pageref{lastpage}}
\maketitle

\begin{abstract}
The Epoch of Reionization (EoR)---when neutral hydrogen in the intergalactic medium was systematically ionized---is a period in our Universe's history that is currently poorly understood. However, a key prediction of most models is a correlation between the density and ionization field during the EoR. This has consequences for the 21cm power spectrum. Here, we propose a parametrization for the density-ionization correlation and study the dependence of the 21cm power spectrum on this parameterization. We use this formalism to forecast the 
ability of current and future observations to constrain these correlations. We find that upper limits on the dimensionless power spectrum at redshifts $7.5 < z < 8.5$ using $k$ bins between $0.1\,\textrm{Mpc}^{-1} < k < 0.75\,\textrm{Mpc}^{-1}$ with error bars at the level of $\sim\!\!20\,\textrm{mK}^2$ about our fiducial model would rule out uncorrelated reionization at $99\%$ credibility. Beyond upper limits, we find that at its full sensitivity, the Hydrogen Epoch of Reionization Array (HERA) will be able to place strong constraints on the sign and magnitude of density-ionization correlations.
\end{abstract}

\begin{keywords}
dark ages, reionization, first stars -- large-scale structure of Universe -- methods: observational -- methods: statistical
\end{keywords}



\section{Introduction}
A key event in our Universe's history is the Epoch of Reionization (EoR) where the neutral hydrogen (HI) making up the intergalactic medium (IGM) is ionized by the first generation of stars and galaxies. The broad timeline of this landmark process is bound by two important measurements: the Cosmic Microwave Background (CMB) and measurements of high redshift quasars at $z < 7$. The CMB suggests that the Universe is neutral by redshift $z \simeq1100$ \citep{Planck} while measurement of the Gunn-Peterson trough at $z \simeq 6$ tells us that the universe must have undergone a transition from neutral to ionized by this time \citep{GPTrough,FanQSO2006,Gallerani2008,McGreer2015}, although recent studies have hinted at a more complicated picture than previously thought \citep{Becker2015,Daloisio2015,Davies2016,Chardin2017,Bosman2018,Becker2018,Eilers2019,Kulkarni2019,Keating2020,Nasir2020}. 
Beyond just its timeline, the astrophysical details of the EoR remain relatively unconstrained observationally. Without direct observational evidence deep into EoR redshifts, the period remains a crucial missing piece in our understanding of galaxy formation.

One of the most promising ways to explore this epoch is to use the hyperfine transition of hydrogen, where 21cm-wavelength photons are absorbed or emitted as electrons in hydrogen atoms flip their spins relative to their protons. (For reviews, see \citealt{Furlanetto2006Review,MoralesWyitheReview,PritchardLoeb2012,LoebFurlanetto2013,LiuShawReview2020}). The 21cm signal allows us to trace primordial hydrogen as a function of redshift and position and can in principle be an incisive probe of first-generation stars and galaxies. The photon lies in the radio part of the electromagnetic spectrum and its absorption or emission is measured relative to the CMB. One therefore measures a \emph{differential} differential brightness temperature $\delta T_b$, which is given by
\begin{eqnarray}
\label{eq:dTb}
 \delta T_b(\mathbf{r}, z) \!\! &\approx& \!\!\! (27\,\textrm{mK}) \left(\frac{T_s(\mathbf{r}, z) - T_\gamma(z)}{T_s(\mathbf{r}, z)}\right)\left[1-x_{\rm HII}(\mathbf{r}, z)\right]\left[1+\delta(\mathbf{r}, z) \right] \nonumber\\
&\phantom{\times} &\times \left [\frac{H(z)/(1+z)}{dv_r/dr}\right ] \left(\frac{1+z}{10}\frac{0.15}{\Omega_mh^2}\right)^{1/2} \left(\frac{\Omega_b h^2}{0.023}\right),
\end{eqnarray}
where $\mathbf{r}$ is a position vector, $z$ is the redshift, $x_{\rm HII}$ is the ionized fraction of hydrogen, $\delta $ is the overdensity, $H(z)$ is the Hubble parameter (with $h$ as its dimensionless counterpart), $T_\gamma(z)$ is the CMB temperature, $\Omega_m$ is the normalized matter density, $\Omega_b$ is the normalized baryon density, and $dv_r/dr$ is the line of sight velocity gradient. The differential brightness temperature depends on the spin temperature $T_s(\mathbf{r}, z)$ of the neutral hydrogen gas, which measures the relative number of HI atoms that are in the excited versus ground hyperfine states. Throughout this paper we consider redshift ranges where the spin temperature is heated above that of the CMB, i.e. $T_s \gg T_{\gamma}$ so that the $T_s$ drops out of Equation \eqref{eq:dTb}, thus allowing us to neglect the spin temperature in our simulations. In this work we set the $\Lambda$CDM parameters to $\sigma_8 = 0.81$, $\Omega_m = 0.31$, $\Omega_b = 0.048$, $h = 0.68$ consistent with Planck 2015 results \citep{Planck}.

Reionization does not occur instantaneously throughout our Universe. Instead, ionized bubbles begin to grow in certain parts of our Universe, eventually coalescing as the process completes \citep{FurlanettoOhPercolation2016}. 
Most models predict that these ionized bubbles are not random; they are tightly correlated with their corresponding density field. There are two extremes in how the ionization field couples to the density field. The ionized bubbles can be positively correlated with the corresponding density field so that the highest density regions correspond to regions of high fractions of ionized hydrogen $x_{\rm HII}$  \citep{Miralda2000,Jordan1}. This is the ``inside-out" model of reionization, where reionization happens first in the most overdense regions before the ionizing photons escape the high density regions and ionize the lower density regions of the IGM.  Conversely, the ionization field can be negatively correlated with the density field. Here regions of the highest density correspond to the lowest fractions of ionized hydrogen and low density regions map to high fractions of ionized hydrogen. This is the ``outside-in" model of reionization. Outside-in models generally require hydrogen recombination rates to dominate the ionizing effects of UV sources so that the most overdense regions have the strongest recombination rates which keep them mostly neutral. In this scenario, reionization happens first in lower density regions where recombination effects are diminished. Outside-in models can also occur if x-rays - which more easily escape high dense regions - play a significant role in ionizing the IGM. In either of these scenarios the highest density regions are last to be ionized. It is also possible for reionization to unfold as a combination of both inside-out and outside-in models \citep{FurlanettoOhCombination2005,MadauHaardt2015}. This scenario entails statistical combinations of the individual inside-out and outside-in statistics between density and ionization fields.

In this paper we build on the work of \citet{WatPrit} and \citet{Binnie}, where the task of distinguishing between inside-out and outside-in scenarios was treated as a discrete model selection problem. Here we take the complementary approach of parametrizing the density-ionization correlation in a continuous manner. This enables us to quantify this correlation, converting the model selection exercise into one of parameter fitting. We can therefore quantitatively constrain outside-in or inside-out morphologies as well as rule out models that predict random uncorrelated bubbles. To this end, we parametrize the correlation between ionization and density field using a single parameter and explore how variation of this parameter affects the 21cm power spectrum. We examine the range of constraints that can be placed on the amount of correlation between these fields using upcoming Hydrogen Epoch of Reionization Experiment (HERA; \citealt{HERA}). In our forecasts, we also include other model parameters that have traditionally been used to capture EoR physics, exploring any new degeneracies that arise from the inclusion of arbitrary density-ionization correlations. This paper therefore serves as a generalization of forecasts such as those of \citet{AdrianMCMC} and \citet{GreigMesinger21cmmc2015}. We do not include as many astrophysical and cosmological parameters as \citet{LiuParsonsForecast2016}, \citet{EwallWiceForecast2016}, \citet{GreigMesingerSimultaneousMCMC2017}, \citet{KernEmulator2017}, or \citet{Park} do, but this is simply to avoid obscuring the discussion of density-ionization correlation, and our formalism can be easily adapted to include any extra parameters that are considered relevant.

The rest of this paper is organized as follows. In Section \ref{sec:Simulations}, we introduce our new correlation parameter and describe how it is incorporated into our simulations. We show how this parameter affects the $21\,\textrm{cm}$ power spectra in Section \ref{sec:Stats}. In Section \ref{sec:HERAforecasts} we describe the HERA instrument and its sensitivity, as well as our forecasting methodology. In Section \ref{sec:MCMCresults} we present the results of our forecasts, emphasizing what one might learn about the morphology of reionization from early observational limits. We summarize our conclusions in Section \ref{sec:Conclusion}.

\section{Simulations}
\label{sec:Simulations}
In inside-out models, ionizing photons are produced in the highest density regions and these photons ionize the surrounding hydrogen before escaping into the lower density regions of the IGM. The correlation between $\delta$ and $x_{\rm HII}$ in inside-out models has consequences for the 21cm brightness temperature because the product $ x_{\rm HII} \delta$ appears in Equation \ref{eq:dTb}. A correlation between density and ionization fields suggests that whenever the overdensity $\delta$ is large, the ionized fraction $x_{\rm HII}$ is large. Therefore the product $x_{\rm HII}\delta$ is typically much larger than the equivalent cross term for outside-in models, leading to lower $\delta T_b$ in general.

In contrast, if recombination rates dominate the reionization process, one ends up with an outside-in scenario. Here, ionizing photons created in high density regions escape and ionize the lower density regions, where recombination rates are lower. As a result, the low density regions remain neutral while the high density regions are kept overall neutral. As reionization proceeds, these high density regions ionize last, and $\delta$ and $x_{\rm HII}$ are negatively correlated. This boosts the brightness temperature contrast, leading to higher $\delta T_b$.

In general, one expects that reionization involves both inside-out and outside-in processes, with the former being important on large scales during early reionization and the latter being important on small scales during late reionization \citep{FurlanettoOhCombination2005}. However, this separation will not necessarily be perfect, and it is possible (for instance) to construct theoretical models where strong, biased sources can result in outside-in effects being important on large scales \citep{Furlanetto2004,Furlanetto2008}. Constraining the sign of the $\delta$-$x_{\rm HII}$ correlation is therefore not just an interesting problem in its own right, but one that may be important in shedding light on the nature of ionizing sources during the EoR \citep{Chang}.  In the subsections that follow, we walk through how one can begin with a semi-numerical simulation of inside-out, map it onto a ``mirror" outside-in simulation, and then finally to generalize to a continuum of simulations with arbitrary density-ionization correlations.
%

\subsection{Inside-out Reionization Simulations Using \texttt{21cmFAST}}
\label{sec:EoR Paramters} 

In order to generate density and ionization fields indicative of a variety of reionization scenarios, we use the publicly available \texttt{21cmFAST} package \citep{21cmFAST}. This semi-numerical code produces an initial density box at high redshift before smoothing the density field to a coarser box corresponding to the same comoving side length. The density field is then evolved through redshift using first order pertubation theory. Throughout our simulations, we use high resolutions boxes of $450^3$ voxels corresponding to a comoving side length of $225\,\textrm{Mpc}$ and coarser boxes of $150^3$ voxels corresponding to the same comoving side length. The code then implements the excursion set formalism of \cite{FZH} for reionization, and predominantly assumes inside-out reionization. This approach evolves the overdensity field $\delta $ as a function of position and smoothing scale and then declares a region ionized if there are enough photons to ionize each baryon within that region of mass $m_{\rm ion}$. These regions must have produced at least enough photons to ionize each of the baryons, satisfying
\begin{equation}
m_{\rm ion} = \zeta m_{\rm gal}
\end{equation}
where $m_{\rm gal}$ is the mass in collapsed objects, and $\zeta$ is the ionizing efficiency of the sources. This condition is continually checked as the density field is smoothed from large scales to small scales. A particular location is flagged as ionized at the first scale in which this condition is satisfied.
The \texttt{21cmFAST} package contains a number of adjustable parameters whose goal is to capture variations in the detailed astrophysics of reionization. We choose three such parameters to vary in this paper: 
\begin{enumerate}
\item The mean free path of ionizing photons, $R_{\rm mfp}$. This is the mean distance an ionizing photon travels before being absorbed by a dense region of hydrogen. This sets the effective horizon for an ionizing photon. We consider values of $R_{\rm mfp}$ between $3$ Mpc and $50$ Mpc.
\item The ionizing efficiency, $\zeta$. The effectiveness of the ionizing sources is summarized by a single parameter 
\begin{equation}
\zeta = \frac{f_{\rm esc}f_*N_\gamma}{(1+n_{\rm rec})},
\end{equation}
which takes into account the amount of ionizing photons produced per stellar baryon $N_\gamma$, the rate of stellar production $f_*$, and the fraction of these ionizing photons that escape into the IGM, $f_{\rm esc}$. The recombination rate, $n_{\rm rec}$, acts to lower the ionizing efficiency. The ionizing efficiency, $\zeta$, is sometimes taken to vary according to halo mass; however, throughout this work, we treat $\zeta$ to be constant during reionization. We consider values of $\zeta$ between $25$ and $50$.
\item The turnover mass, $M_{\rm turn}$. This describes the mass of a halo below which there is an exponential suppression in star formation efficiency. A turnover mass of $10^8 M_{\odot}$ corresponds roughly to $\sim10^4\,\textrm{K}$ virial temperature, at which point gas in a halo can start to cool via Ly$\alpha$ emission. We consider values of $M_{\rm turn}$ between $10^7 M_{\odot}$ and $10^9 M_{\odot}$.
\end{enumerate}

Variation of these parameters modify the onset and duration of reionizaton, and for fixed redshift can also change the size of the ionized bubbles. Models with large values of $\zeta$ imply that emitters are very effective at ionizing the IGM and increasing this parameter shifts the onset of reionization to higher redshifts. The size and scale of the bubbles are also dependent on the turnover mass. Smaller values of $M_{\rm turn}$ allow for smaller sources to contribute to the EoR. As a result, ionized bubbles tend to be smaller and more numerous. Reionization is complete when these bubbles merge together to form a complete ionized IGM. Modification of $R_{\rm mfp}$ also sets the maximum size of the ionized bubbles which can have consequences for large scales as the ionized bubbles grow large enough to approach the maximum size $R_{\rm mfp}$. Variation of these parameters have consequences for the structure of the temperature field $\delta T_b$. However, these variations are still within the context of a predominantly inside-out formalism, with the ionized fraction field $x_{\rm HII}$ generally correlated with the overdensity field $\delta$. We therefore now need to modify the prescription by which the ionized fraction field (or equivalently, the ionization field) is generated from an overdensity field.


\subsection{Outside-in Reionzation Simulations using \texttt{21cmFAST}}
\label{sec:signflip}

In order to make outside-in temperature maps using \texttt{21cmFAST} we require that the temperature maps be produced from density and ionized fraction fields that are negatively correlated. To do so starting from an inside-out simulation package, we adapt methods from the structure formation literature and use the idea of sign-flipped initial conditions \citep{PontzenPairedSimulations2016,Paco2018}. In particular, we can produce $\delta$ and $x_{\rm HII}$ fields that are negatively correlated with one another by pairing a sign-flipped overdensity field with its original ionization field. A sign flip will invert $\delta$ while preserving the rest of its structure. Importantly, this sign flip must be applied to the initial (i.e., high-redshift) density field while $\delta$ is still in the linear regime. This ensures that the sign flipped density and velocity fields have undergone the correct non-linear evolution, resulting in physically plausible probability distribution functions for both quantities. The evolved sign flipped density field will contain the same statistics as the original density field except overdense regions become underdense regions and vice versa. When $\delta^{\rm flip}$ is paired with its original ionization field $x_{\rm HII}$, the regions of lowest baryonic overdensities now correspond to regions of high ionized fractions of hydrogen in the ionization field. Similarly, regions of high baryonic overdensites in $\delta^{\rm flip}$ correspond to neutral regions of hydrogen in $x_{\rm HII}$.  The fields $x_{\rm HII}$ and $\delta^{\rm flip}_b$ are therefore anti-correlated. Temperature maps computed using Equation \eqref{eq:dTb} with fields $\delta^{\rm flip}_b$ and $x_{\rm HII}$ will have statistics representative of outside-in reionization.

\begin{figure*}
    \includegraphics[width=15cm]{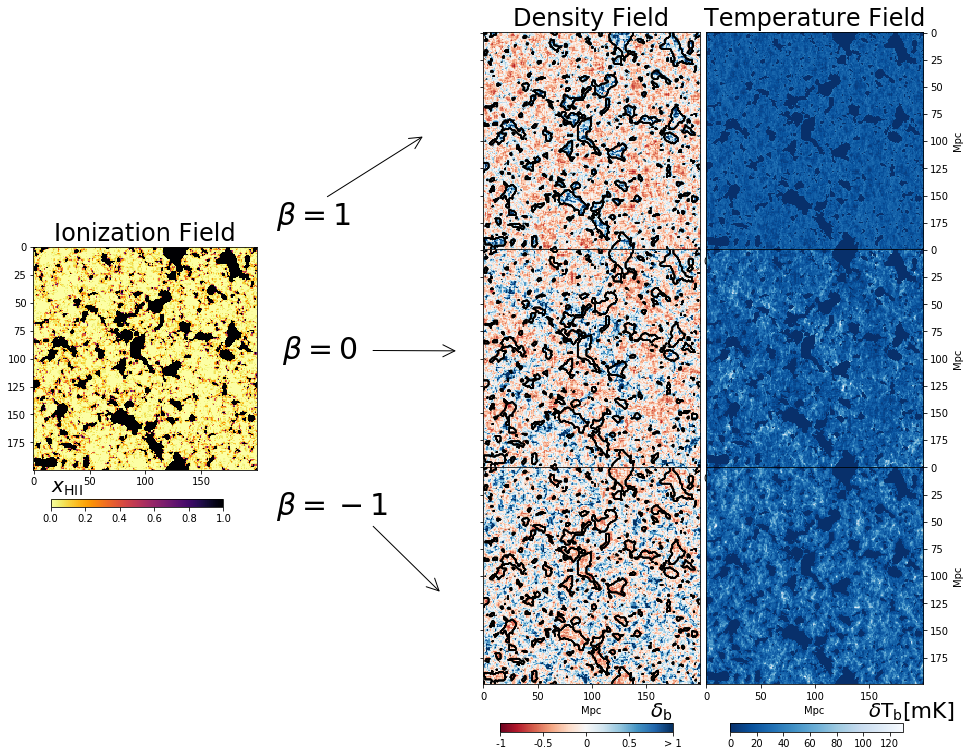}
    \caption{Example ionization $x_{\rm HII}$, density $\delta$, and brightness temperature $\delta T_b$ fields at $z=9$ with fiducial parameters of $\zeta_0= 25$  , $M_{\textrm{turn}, 0} = 5\times10^8 M_\odot $, and $R_{\textrm{mfp}, 0} = 30\,\textrm{Mpc}$. The ionized regions from the ionization field are superimposed as contours on the density field to emphasize the correlation between the two fields. Inside-out scenarios (with $\beta > 0$) have positively correlated density and ionization fields. Conversely, outside-in scenarios (with $\beta < 0$) have negatively correlated fields. The case $\beta = 0$ refers to the scenario where the ionization field and density field are entirely uncorrelated. The $21\,\textrm{cm}$ brightness temperature field is sensitive to the type of correlation between matter and ionization fields. For example, the outside-in model shown here has the highest brightness temperature, since regions of high density are also regions of highest neutral fraction.}
    \label{fig:tempmaps}
\end{figure*}

In Figure \ref{fig:tempmaps}, we provide some visual intuition for our inside-out and outside-in models. The leftmost picture is of an ionization field that can be paired with density fields (middle column) that are either anticorrelated (bottom row), uncorrelated (middle row), or correlated (top row). Contours around ionized bubbles from the ionization field are superimposed on each density field. The positively correlated density field is the original density field from which \texttt{21cmFAST} produced the ionization field; it therefore represents an inside-out scenario, and one sees that regions of high density correspond to regions of high ionization. The negatively correlated density field comes from the sign-flipped method, where high density regions correspond to regions of high neutral fraction; this represents an outside-in scenario. The $x_\textrm{HII} \delta $ term of Equation \ref{eq:dTb} then predicts a lower brightness temperature for inside-out models than for outside-in models, as is visually confirmed in maps of $\delta T_b$ (right column). 

 The morphologies for our outside-in maps differ from outside-in maps that model the sub-grid physics. For example, \cite{Jordan1} produce outside-in maps by modeling the inhomogenous recombinations that occur near overdense regions and then parametrize the degree to which recombinations affect reionization by tuning the recombination timescale. This approach produces outside-in maps which are driven by self shielded ovedense regions. In this scenario, reionization begins as inside-out but the morphology changes after the midpoint of reionization when the shielded high-density neutral regions ionize the IGM. These overdense neutral regions are found inside the ionized bubbles and increase the small scale structure while also limiting the size of coherently ionized regions. This decreases the variance on large scales (i.e., the power spectrum at low $k$; see Section \ref{sec:Stats}) as compared to our model. We find that the variance on all scales between these two approaches diverge as reionization progresses. Thus, unlike the very physically motivated model such as that of \cite{Jordan1}, our outside-in model should be viewed more as a phenomenological parametrization to bracket the possibilities. Our approach is similar to that of \cite{WatPrit} which produces an outside-in model by implementing an inversion operation on a fiducial inside-out model. Should early data favour outside-in reionization, our framework is general enough that one could adopt a physically motivated outside-in model and to decorrelate the density field from there. We are mostly interested in the correlation statistics indicative of these reionization morphologies in order to rule out uncorrelated reionization scenarios.
 
 Note that our procedure for creating outside-in maps from inside-out maps changes the mass-weighted ionization history. This occurs because the sign flip acts on the density field and not the ionization field. By keeping the ionization field the same, the total volume of the ionized regions remains constant, but these regions are now paired with different density values thereby changing the mass weighted global ionized fraction.  The largest difference between the mass-weighted ionization histories of these extreme models is about $\pm 0.1$, occurring near the midpoint of reionization. This does not constitute a significant difference in our results. 

\subsection{Simulating Arbitrary Correlations using \texttt{21cmFAST}}
\label{sec:DecorrProcedure}
It is also possible to generate temperature fields from $x_{\rm HII}$ and $\delta$ where $\delta$ and $x_{\rm HII}$ are correlated by some arbitrary amount. To do this, we draw a random phase $\phi $ from a Gaussian of standard deviation $\sigma$ and apply this phase constant to each point of the Fourier transformed density field $\widetilde{\delta}$ in $k$ space, such that $\widetilde{\delta} (k)\rightarrow \widetilde{\delta}(k) e^{i\phi}$. This shifts the phase of each Fourier mode while leaving the overall variance of the field for each Fourier mode unchanged. Just as with the sign-flipped overdensity box, this ensures that the statistics of the density field (in particular, the power spectrum---see Section \ref{sec:Stats}) remain unchanged. However, upon returning the overdensity box to configuration space, high- and low-density regions will have shifted from their original positions, somewhat decorrelating the field from its corresponding neutral fraction box. This procedure is carried out on the fine resolution boxes of 450$^3$ voxels. This allows for the smallest Fourier modes to be decorrelated before they are smoothed over to the coarser resolution box. Our results in Section \ref{sec:Stats} do not depend on the resolution of the fine resolution box---so long as they fall within the acceptable range of resolutions set by \texttt{21cmFAST}.

The degree of correlation between the ionization field and the density field is governed by $\sigma$. If $\sigma $ is small, then the randomly chosen values $\phi $ are also typically small and so the Fourier modes are not significantly shifted. Once the field is transformed back to configuration space, the resulting density field only be slightly perturbed. As $\sigma $ is increased, $\phi$ are chosen from an increasingly broad range, and so can produce large deviations from their original configurations. Upon transformation to configuration space, the density field becomes increasingly decorrelated from its original ionization field. As $\sigma$ approaches $\pi$, the multiplicative factor $e^{i\phi}$ essentially randomizes the phases, and the resulting density field is completely uncorrelated. The quantity $\sigma $ therefore quantifies the amount that the density field has been decorrelated from the $x_{\rm HII}$ field: for the special case of  $\sigma = 0 $ we recover the original density field while for $\sigma \sim  \pi $ the density field is entirely decoupled from $x_{\rm HII}$. Identical applications of a given value of $\sigma$ with different random seeds lead to slight different realizations of the density field. However the fluctuations in the resulting statistics (defined in Section \ref{sec:Stats}) are insignificant (a fractional shift of $10^{-4}$ in the variance of the field) and do not impact our results in Section \ref{sec:MCMCresults}.

If the above decorrelation method is combined with the sign flip from Section \ref{sec:signflip}, we can also introduce arbitrary levels of anticorrelation between $\delta$ and $x_{\rm HII}$ for outside-in models of reionization. The sign flip and decorrelation can then be folded into a single parameter $\beta$, which is defined as
\begin{equation}
\beta \equiv 
\begin{cases}
\textrm{sgn}(\sigma) \left( 1 - \frac{|\sigma|}{\pi} \right) & \sigma \neq 0 \\
\pm 1 & \sigma =0
\end{cases}
\label{eq:betaparam}
\end{equation}
%
where as shorthand, we use the sign $\textrm{sgn}(\sigma)$ of $\sigma$ to record whether we are decorrelating from an inside-out model ($\textrm{sgn}(\sigma)= +1$) or an outside-in model ($\textrm{sgn}(\sigma)= -1$). The special case $\sigma = 0$ is multivalued in $\beta$, but since those correspond to the original inside-out and outside-in models, we simply assign those to $\beta = +1$ and $\beta = -1$ respectively. With our definition of $\beta$, we therefore have a parameter that can be continuously dialled from $+1$ to $-1$ to go from a fully inside-out scenario to a fully outside-in scenario: a positive value of $\beta$ indicates a scenario where an initially correlated matter and ionization field are decorrelated by $\sigma$ while a negative $\beta$ indicates a scenario where a negatively correlated matter and ionization field are decorrelated by $\sigma$.
Table \ref{tab:amps1} summarizes the terminology used to describe the type of correlation as well as the model it pertains to. In the sections that follow, we will consider $\beta$ to be another one of our EoR parameters (joining $\zeta$, $R_{\rm mfp}$ or $M_{\rm turn}$) that can be potentially constrained by $21\,\textrm{cm}$ observations.%


\begin{table}
\caption{Lexicon for physical models and their respective correlations \label{tab:amps1}}
\begin{center}
\begin{tabular}{|c|c|c|} 
\hline
$\beta$ & Moniker for Field correlations  & Physical Model \\ 
        &   $x_{\rm HII}$ $\delta $   &               \\
\hline\hline
1 &  Correlated &  Inside-out\\
\hline
  $1 < \beta < 0$ & Increasingly correlated & Mostly inside-out\\ 
\hline
 $0$  & Uncorrelated & Uncorrelated\\ 
  \hline
 $0 < \beta < -1$ & Increasingly anti-correlated & Mostly outside-in\\ 
\hline
  $-1$ & Anti-correlated & Outside-in\\ 
  \hline
\end{tabular}
\end{center}
\end{table}

\section{Temperature Field Statistics}
\label{sec:Stats}

\begin{figure}
    \includegraphics[width=8cm]{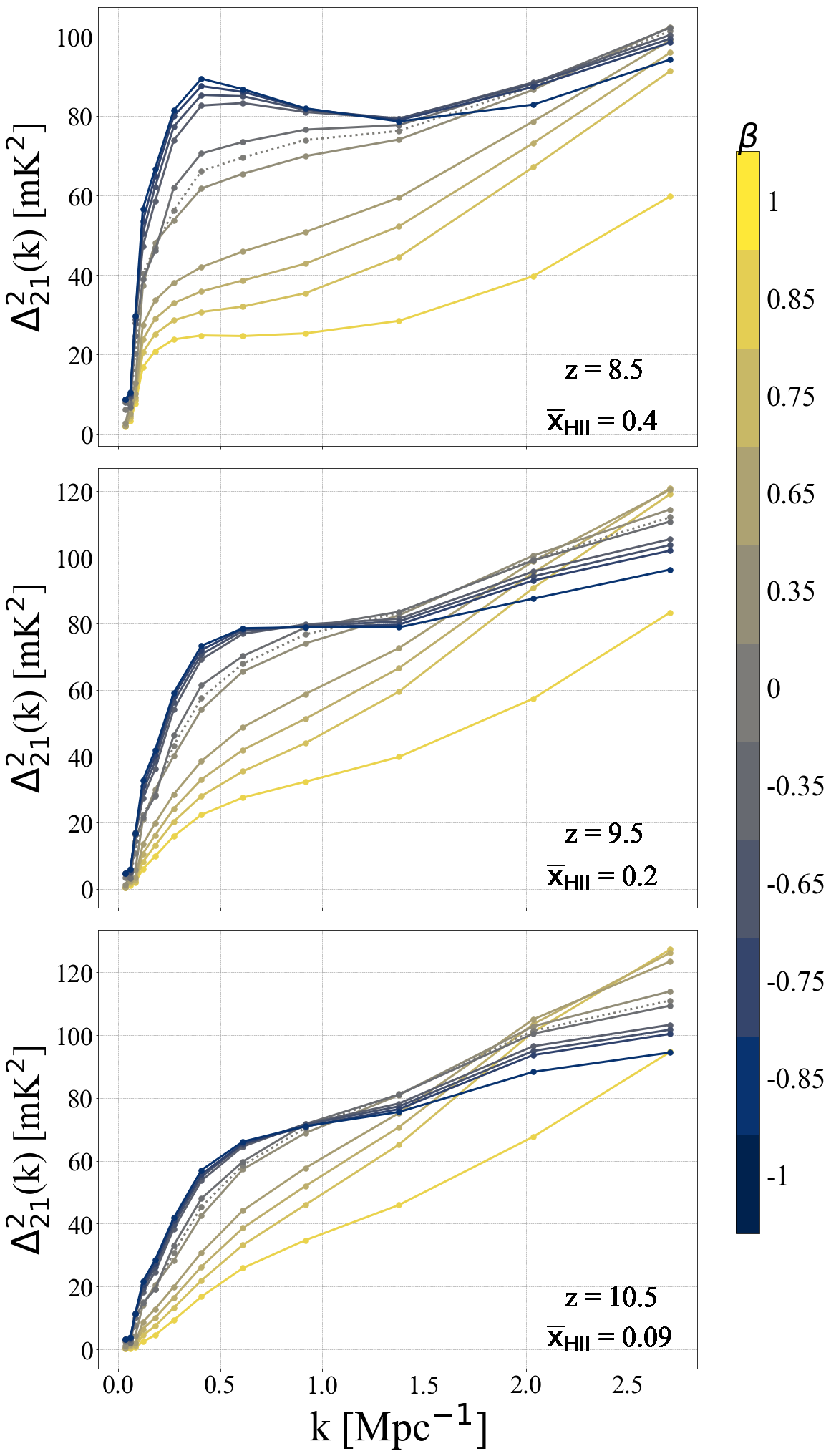}
    \caption{Power Spectra for varying $\beta$ at different redshifts, with the other EoR parameters fixed to fiducial values of $\zeta_0$ = 25 , $M_{\textrm{turn}, 0} = 5\times10^8 M_\odot $, and $R_{\textrm{mrp}, 0} = 30\,\textrm{Mpc}$. The mean ionized fractions $\overline{x}_{\rm HII}$ at redshifts, $z = 10.5$, $z = 8.5$ are 0.09, 0.2 and 0.4 respectively. The amplitude of $\Delta^2_{21}$ increases as $\beta$ is decreased from its maximal value $\beta = 1$, corresponding to inside-out models of reionization. The amplitude of the power is maximum at $\beta = - 1$, corresponding to outside-in models. The contrast between these models is largest at the scale corresponding to the size of the ionized regions. The dotted line represents a model without any correlation between density and ionization fields.}
    \label{fig:vert_pspec}
\end{figure}

In Section 2 we gained intuition for how correlations between ionization and density fields can affect the 21cm temperature maps. In this section we compute the statistical properties of these maps that can be measured by upcoming 21cm instruments. In particular, we compute the ``dimensionless" 21cm power spectrum $\Delta^2_{21}$ which is defined through the brightness temperature field as
\begin{equation}
    \Delta^2_{21}(k) \equiv \frac{k^3}{2\pi^2} \frac{ \langle | \widetilde{\delta T_b }(\mathbf{k}) |^2\rangle }{V}
\end{equation}

where $V$ is the survey volume, $\widetilde{\delta T_b}$ is the Fourier transform of the brightness temperature field (into a space defined by spatial wavevector $\mathbf{k}$), and the angular brackets indicate an average over shells of constant $k\equiv |\mathbf{k}|$. Physically, $\Delta^2_{21}$ measures the contribution to the brightness temperature variance per logarithmic interval in $k$. The brightness temperature is sensitive to the inside-out versus outside-in morphology via the $x_\textrm{HII} \delta$ cross term in Equation \eqref{eq:dTb}. Importantly, notice that the power spectrum does not depend on the phases of the complex Fourier field. Defining the analogous quantity for the density field (i.e., defining the matter power spectrum) then immediately reveals that our decorrelation procedure from Section \ref{sec:DecorrProcedure}---which only involved altering the Fourier phases of our density field---preserves the statistics of the matter distribution. Each decorrelated field is therefore a perfectly legitimate realization of $\delta$.
%

\begin{figure*}
\centering
\begin{minipage}{0.5\textwidth}
  \centering
  \includegraphics[width=.95\textwidth]{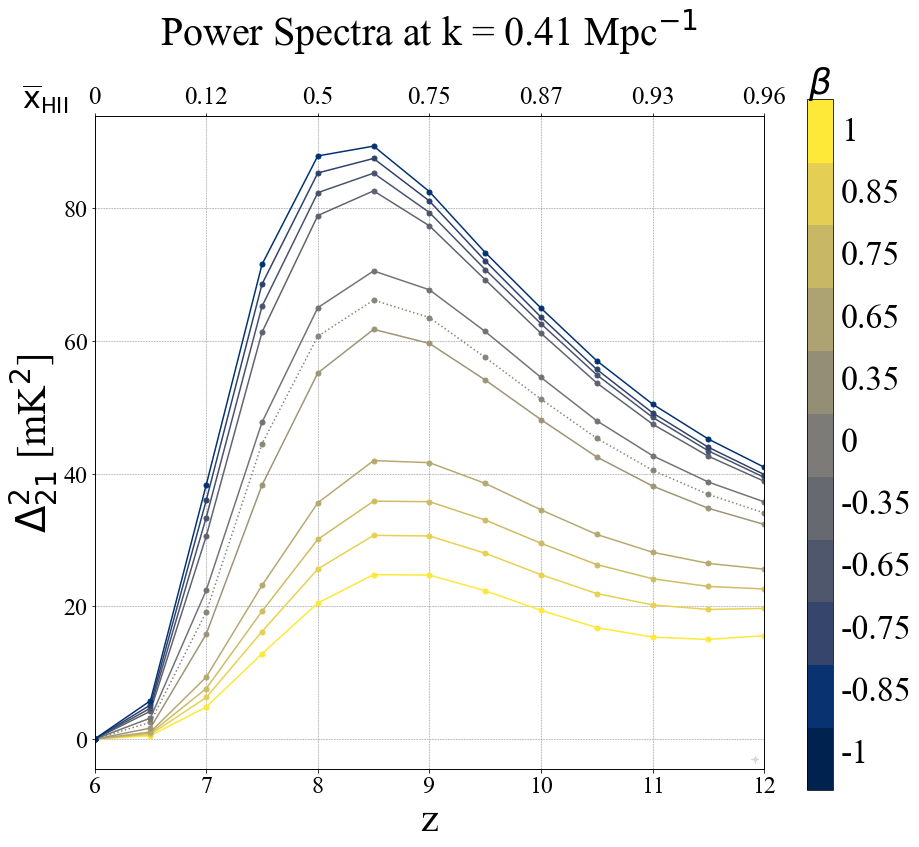}
\end{minipage}%
\begin{minipage}{0.5\textwidth}
  \centering
  \includegraphics[width=0.95\textwidth]{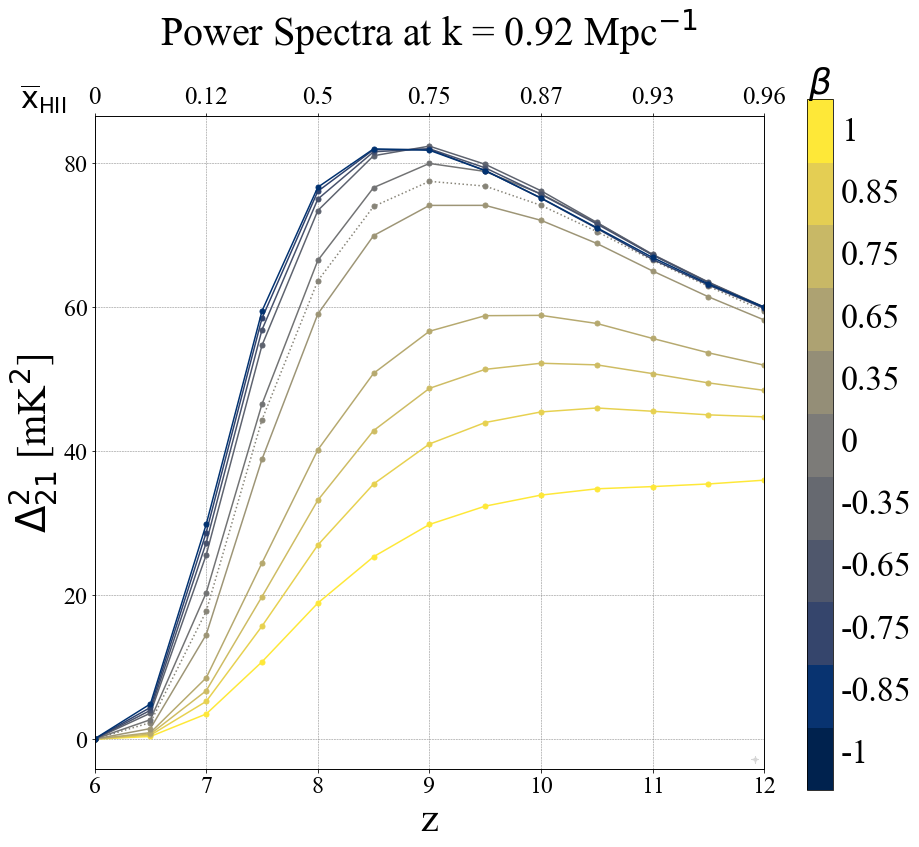}
\end{minipage}
\caption{Power spectra as a function of $z$ and $\beta$ at two chosen $k$ values. The contrast between outside-in and inside-out models is largest at the midpoint of reionization. For the fiducial parameters of $\zeta_0$ = 25 , $M_{\textrm{turn}, 0} = 5\times10^8 M_\odot $, and $R_{\textrm{mfp}, 0} = 30\,\textrm{Mpc}$, this occurs at $z \approx 8$. The dotted curve represents a model without any correlations between $\delta$ and $xH_{\rm II}$. }
\label{fig:z_evolution}
\end{figure*}

\subsection{Variation of $\Delta^2_{ \rm 21}$ as a Function of $\beta$}
Figure \ref{fig:vert_pspec} illustrates the effect of varying $\beta$ on the power spectrum at three different redshifts. 
Beginning from an inside-out ($\beta = 1$) model, decreasing $\beta$ reduces the density field's original correlation with $x_{\rm HII}$, and increases the chances that neutral regions overlap with overdense regions in $\delta$. As a result, we find increasing power on large scales as we decrease $\beta$ from $+1$ to $-1$. The intermediate scenario with $\beta = 0$ is one where the density field and the ionization field are uncorrelated, and as expected, we find that the predictions for $\Delta^2_{21}$ in this case are the same whether we decorrelate from an initially inside-out model or an outside-in model (i.e., whether we approach $\beta =0$ from below or above). We also find that the qualitative behaviour of Figure \ref{fig:vert_pspec} is insensitive to the exact form of our $\beta$ parametrization, Equation \eqref{eq:betaparam}. Modifying the parameterization changes the precise rate with which $\Delta^2_{21}$ changes as a function of $\beta$, but the extreme scenarios remain the same and are given by the $\beta = \pm 1$ curves of Figure \ref{fig:vert_pspec}.

\begin{figure}
  \includegraphics[width=0.49\textwidth]{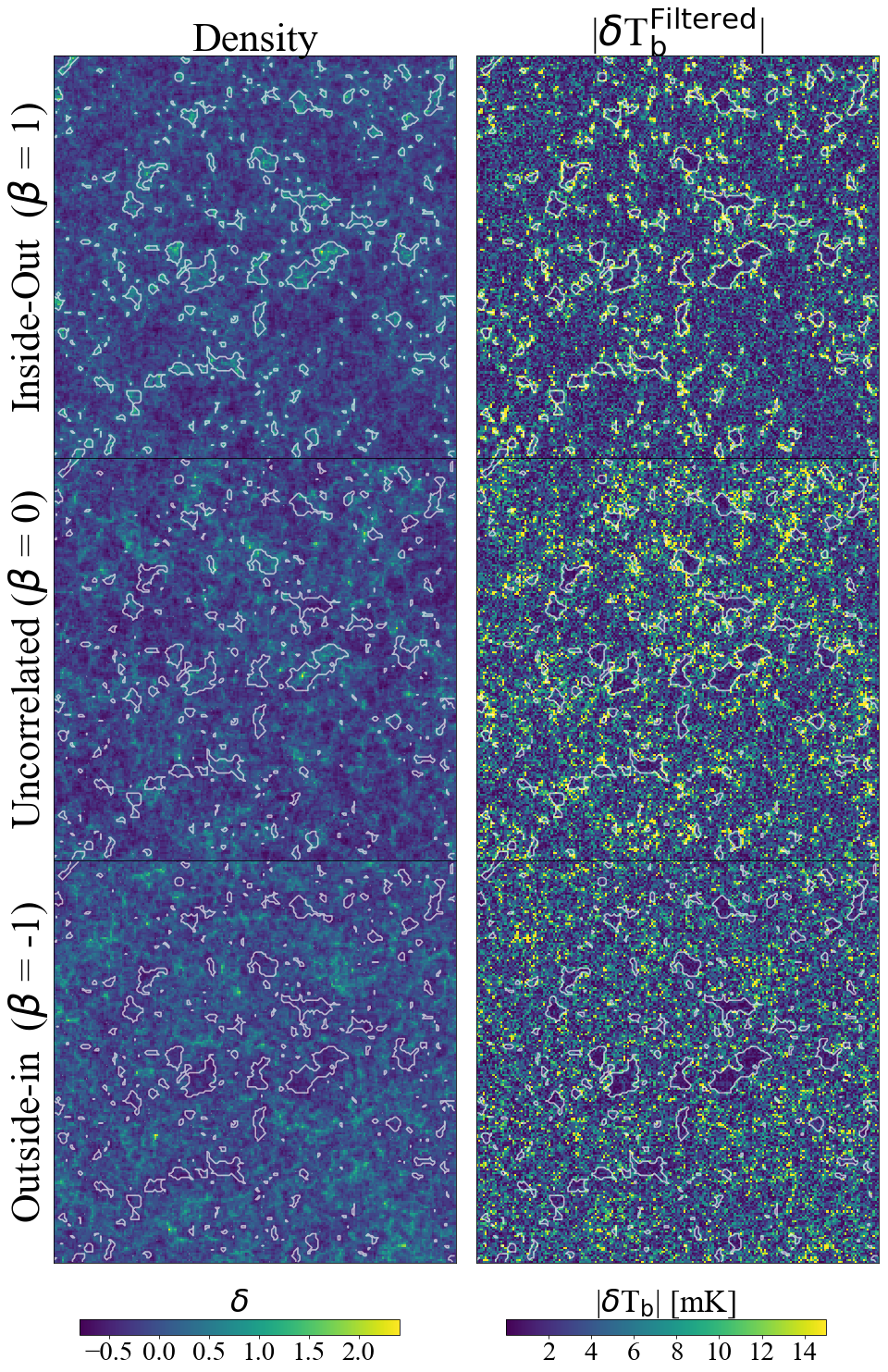}
  \caption{Density boxes (left) alongside their resulting temperature fields (right), with contours showing the locations of ionized bubbles overlaid. Fourier modes with $k < 2\,$Mpc$^{-1}$ have been filtered out from $\delta $T$_{\rm b}$ to elucidate the high $k$ behavior seen in Figure \ref{fig:vert_pspec}. In this high-$k$ regime, large amplitude brightness fluctuations are seen throughout the neutral regions }
\label{fig:filtered_boxes}
\end{figure}

One curious feature seen in Figure \ref{fig:vert_pspec} is the fact that the behaviour in $\Delta^2_{21}$ is non-monotonic with $\beta$ at high $k$. That is, at high $k$ we see that it is in fact the uncorrelated case with $\beta = 0$ that has the highest power, with the extreme cases of $\beta = \pm 1$ having lower power at $k > 1.5 \,\textrm{Mpc}^{-1}$. To get some intuition for why this is the case, consider Figure \ref{fig:filtered_boxes}, where we show high-pass filtered versions of the brightness temperature field  for $\beta = +1$, $0$, and $-1$ at $z=9$. These filtered temperature fields contain only power at $k > 1.5\,\textrm{Mpc}^{-1}$, and for clarity (to bring out the fluctuations) we plot their absolute values. Also plotted are the density fields for comparison, and overlaid on both fields are contours demarcating the ionized bubbles. What one sees is that as expected, for the extreme inside-out ($\beta = +1$) or outside-in ($\beta = -1$) cases, the locations of high density regions are dictated by the ionized bubbles. As before, the inside-out case places its highest density peaks inside the ionized bubbles, ``squandering" the opportunity to achieve high brightness temperatures. The outside-in case places its highest density peaks in neutral regions. But the fact that it preferentially places its peaks away from bubbles means that there is less volume available in this scenario to produce high brightness temperature spots. In contrast, the uncorrelated scenario indiscriminately places bright spots throughout the volume, thus utilizing more of it to give a large integrated signal.

Our interpretation of this high-$k$ behaviour is bolstered by the redshift dependence seen in Figure \ref{fig:vert_pspec}. The non-monotonic behaviour with $\beta$ is strongest during the earliest phases of reionization, where most of the universe is neutral. In such a regime, not utilizing the full volume results in a greater reduction in the fluctuation power. As reionization proceeds, a random indiscriminate distribution of density peaks (as is the case with $\beta = 0$) becomes more and more disadvantageous as these peaks become more and more likely to fall into an ionized bubble. The outside-in model thus begins to get brighter relative to the uncorrelated model. Indeed, we find that after the midpoint of reionization ($z \sim 8$ for our fiducial simulations), when there is less neutral volume remaining, the trends in the power spectra are once again monotonic in $\beta$.

Although the behaviour at high $k$ is interesting, it has essentially no impact on the forecasts that we will present in Section \ref{sec:MCMCresults}, given that instrument sensitivities tend to be low in those regimes. However, for future-generation instruments, the high-$k$ trend may be a signature that is worth pursuing, as it may break some of the parameter degeneracies that we will find in Section \ref{sec:MCMCresults}.


\subsection{Redshift Evolution of the Signal}

Focusing now on the low-$k$ regime that drives our forecasts, Figure \ref{fig:z_evolution} shows how $\Delta^2_{21}$ evolves as a function of $z$ and $\beta$. 
During the first half of reionization, the ionized bubbles are still small and so reionization has yet to make a significant imprint on the brightness temperature field. Altering the density-ionization correlation via $\beta$ thus has little effect on the power spectrum and all the curves converge at high $z$. As one approaches a global ionization fraction of $\sim 0.5$, the ionized bubble morphology has its largest influence on the power spectrum, and thus it is there that one sees the greatest sensitivity to $\beta$.
As one approaches the end of reionization, the 21cm signal vanishes and $\Delta^2_{21}$ once again loses its sensitivity to $\beta$. 

The redshift space distortions contained in the temperature maps also affect $\Delta^2_{21}$. Redshift space distortions induce an angular structure to $P(\mathbf{k})$, but also boost the power evenly in the angularly averaged $P(k)$ by a geometric factor of $1.87$ \citep{BarkanaLoeb}. We find that they primarily affect outside-in maps and not their inside-out counterparts. The effect of redshift space distortions on $\delta T_b$ are greatest near high density regions, which are first to be ionized in an inside-out scenario. Therefore, the effect of redshift space distortions on $\Delta^2_{\rm 21}$ is small for redshifts $z < 12$ \citep{21cmFAST} in the inside-out scenario. Conversely, the high density regions are ionized last in outside-in maps allowing for a more impactful effect on $\Delta^2_{21}$ throughout reionization. We find a maximum boost of $15\,\textrm{mK}^2$ in $\Delta^2_{\rm 21}$ due to redshift space distortions occurring in the extreme outside-in scenarios of $\beta = -1$ near the midpoint of reionization. The effects on the corresponding inside-out case is negligible within the same regime. In producing $\Delta^2_{\rm 21}$, we take a more conservative approach and average over the angular dependence for all $\beta$ realisations.The results presented in Figure \ref{fig:vert_pspec} include these effects. We take this conservative approach in producing our forecasts in Section \ref{sec:MCMCresults}. This approach is more appropriate for early limits, where experiments are not expected to have a high enough signal-to-noise per Fourier mode for characterizing the angular patterns in $\mathbf{k}$ needed to easily measure redshift space distortions. In the future, however, one could imagine using redshift space distortions as another avenue for constraining density-ionization correlations in a more extended analysis.



In addition to the sensitivity of $\Delta^2_{\rm 21}$ on $\beta$, the other EoR parameters described in Section 1 will also affect the power spectrum. We want to explore the extent to which one can uniquely constrain $\beta$ from measurements of $\Delta^2_{\rm 21}$. We perform these numerical forecasts in the following sections.
\section{HERA Forecasts}
\label{sec:HERAforecasts}

We now bring together the parametrization and intuition of Section \ref{sec:Simulations} and the power spectrum predictions of Section \ref{sec:Stats} to provide forecasts on potential observational constraints on our correlation parameter $\beta$. In this section we lay out our forecasting methodology, reserving a discussion of our results for Section \ref{sec:MCMCresults}.

\subsection{Fiducial Instrument and Sensitivity}
We use HERA as the fiducial instrument for our forecasts, although similar results can be easily obtained for other current and upcoming instruments such as the Low Frequency Array \citep{LOFAR2013}, the Murchison Widefield Array \citep{Bowman2013,Tingay2013}, and the Square Kilometre Array \citep{SKAI}. When completed, HERA will consist of 350 parabolic dishes, each $14\,\textrm{m}$ in diameter observing from $50\,\textrm{MHz}$ to $250\,\textrm{MHz}$. Its forecasted sensitivity is such that $> 20\sigma$ detections of the $21\,\textrm{cm}$ power spectrum from the EoR should be possible \citep{AdrianMCMC,HERA}.

%

To model HERA's expected error bars on a power spectrum measurement, we make use of the publicly available code \texttt{21cmSense} \citep{PoberBAOBAB2013,AdrianMCMC}. At a given instant, a particular pair of antennas of the interferometer separated by a baseline vector $\mathbf{b}$ approximately samples a Fourier mode of the sky with a wavevector $\mathbf{k}_\perp$ oriented perpendicular to the line-of-sight. The sampled wavevector is given by $\mathbf{k}_\perp \approx 2 \pi \mathbf{b} / \lambda X$, where $\lambda$ is the observation wavelength and $X$ is a conversion factor from angular separation $ \theta$ to transverse comoving distance $r_\perp$, and is given by
\begin{equation}
    X  \equiv \frac{r_\perp}{\theta} = \frac{c}{H_0}\int^z_0 \frac{dz'}{E(z')}
\end{equation}
with $c$ the speed of light, $H_0$ the Hubble parameter today, $E(z) \equiv \sqrt{\Omega_m(1+z)^3 + \Omega_\Lambda}$ and $\Omega_\Lambda$ the normalized dark energy density. Fourier modes along the line of sight are probed using the frequency spectrum, since different observed frequencies correspond to different redshifts, with a conversion factor $Y$ converting between frequency interval $\Delta \nu$ and increments in radial comoving distances $\Delta r_\parallel$:
\begin{equation}
    Y \equiv \frac{\Delta r_\parallel}{\Delta \nu} = \frac{c}{H_0 \nu_{21}} \frac{(1+z)^2}{E(z)},
\end{equation}
where $\nu_{21} \approx 1420\,\textrm{MHz}$ is the rest frequency of the $21\,\textrm{cm}$ line. The \texttt{21cmSense} package computes the amount of time $t_\textrm{int}$ that an interferometer spends observing a particular mode in $k_\perp$-$k_\parallel$ space and assigns an error bar $\varepsilon$ to a hypothetical power spectrum measurement, where 
\begin{equation}
\label{eq:sensitivities}
         \varepsilon( { k} ) = X^2 Y \frac{{k^3} \Omega_{\rm eff} }{2 \pi^2 } \frac{T^2_{\rm sys}}{2 t_{\rm int}}
\end{equation}
where $T_\textrm{sys}$ is the system temperature of the telescope (generally dominated by the brightness temperature of the sky at these frequencies), and $\Omega_{\rm eff}$ is the effective solid angle of the primary beam of each dish \citep{ParsonsLimit2014}.

Beyond signal-to-noise considerations, a realistic forecast must account for foreground contaminants. Astrophysical---but non-cosmological---sources of emission are bright in the low-frequency radio spectrum, and these foregrounds are expected to be brighter than the cosmological $21\,\textrm{cm}$ signal by many orders of magnitude in brightness temperature. Fortunately, these foregrounds are expected to be spectrally smooth, and thus they predominantly contaminate a select triangular region in Fourier space known as ``the wedge" that is given by
\begin{equation}
\label{eq:wedge}
k_\parallel \leq \left( \frac{X}{\nu Y}\right) k_\perp,
\end{equation}
where $\nu$ is the observation frequency. (For a derivation and a discussion of various subtleties associated with this equation, see \citealt{LiuShawReview2020} and references therein). In our forecasts, we employ the ``moderate" foreground setting in \texttt{21cmSense}, which simply states that modes satisfying Equation \eqref{eq:wedge}, plus those that are up to $0.1\,h\textrm{Mpc}^{-1}$ higher in $k_\parallel$, are considered irretrievably contaminated by foregrounds and are discarded in one's analysis. The extra buffer of $0.1\,h\textrm{Mpc}^{-1}$ accounts for the possibility that low levels of spectral unsmoothness in one's foregrounds may cause a bleed to higher $k_\parallel$ than one might theoretically expect \citep{21cmsense1}.In generating the full error bars for our forecasts, we add sample variance to the telescope sensitivities computed in Equation \ref{eq:sensitivities}. The sample variance is generated using a fiducial EoR inside-out model. Using a fiducial outside-in or uncorrelated EoR model to generate the sample variance does not qualitatively change the results in Section \ref{sec:MCMCresults}.


 \subsection{Markov Chain Monte Carlo Setup}
\label{sec:MCMCsetup}

With a model for power spectrum sensitivities, performing a forecast is tantamount to evaluating the posterior probability distribution for the parameters of interest. In particular, suppose we measure a power spectrum $\Delta^2_{21} (k,z)$ from HERA and group all the measurements from different $k$ and $z$ bins into a data vector $\mathbf{d}$. Grouping our model parameters $\beta$, $M_{\rm turn}$, $R_{\rm mfp}$, and $\zeta$ into another vector $\boldsymbol \theta$, our goal is to use Bayes' theorem to find the posterior $p(\boldsymbol \theta | \mathbf{d})$, i.e., 
\begin{equation}
p(\boldsymbol \theta | \mathbf{d}) \propto p( \mathbf{d} | \boldsymbol \theta ) p(\boldsymbol \theta),
\end{equation}
where $p( \mathbf{d} | \boldsymbol \theta ) $ is the likelihood function and $p(\boldsymbol \theta)$ is our prior.
In our forecasts, we impose uniform priors on all parameters. For $\zeta$ we pick $10 < \zeta < 100$ which is broadly consistent with CMB and Ly$\alpha$ constraints on reionization \citep{kSz}. For $R_{\rm mfp}$ we say that $3\,\textrm{Mpc} < R_{ \rm mfp} < 80\,\textrm{Mpc}$ to span a reasonable range in uncertainty on the parameter \citep{Rmfp}. For $M_{\rm turn}$ we use $10^7 M_\odot <  M_{\rm turn} < 9 \times 10^9 M_\odot$. This is motivated by the atomic cooling threshold and by current constraints on the faint end of UV luminosity functions \citep{Park}. Finally, for $\beta$ we adopt a uniform prior with $-1 \leq \beta \leq 1$, which spans the full range of density-ionization correlations discussed in Section \ref{sec:Simulations}.
%

The likelihood $p( \mathbf{d} | \boldsymbol \theta )$ is non-analytic in the EoR parameters. To compute it, we generate model predictions for the density and ionization fields from \texttt{21cmFAST} simulations, given a combination of $\zeta$, $R_{\rm mfp}$, and $M_{\rm turn}$. We then regenerate the density field as appropriate for the desired level of decorrelation from the ionization field as specified by the $\beta$ parameter. This updated density field is then used in conjunction with the original ionization field to form a brightness temperature field using Equation \eqref{eq:dTb}. The power spectrum of the resulting maps $\Delta^2_{\rm model}$ are then computed and compared to the ``measured" power spectrum $\Delta^2_{21} (k,z)$ via a Gaussian likelihood of the form
 through the chi squared $\chi^2$ statistic given by
\begin{equation}
    p( \mathbf{d} | \boldsymbol \theta ) \propto \exp \left[-\frac{1}{2} \sum_{z , k} \frac{ \left(\Delta^2_{\rm model} - \Delta^2_{21} \right)^2}{\varepsilon^2} \right],
\end{equation} 
where we have assumed that all the $k$ and $z$ bins are statistically independent. Our forecasts consider different combinations (see Section \ref{sec:MCMCresults}) of redshifts $ z= 6$ to $z = 10$ in steps of $\Delta z = 0.5$, with an observational bandwidth $\Delta \nu \equiv \nu_{21} \Delta z/(1+z)^2\Delta z$ for each redshift bin. 
We exclude bins $ k > 0.75\,\textrm{Mpc}^{-1}$ for computational simplicity as the HERA error bars are large in that regime and including those $k$ bins do not add alter our forecasts significantly.

To sample our posterior distribution, we use a Markov Chain Monte Carlo (MCMC) approach, as implemented by the affine invariant MCMC package \texttt{emcee} \citep{emcee}. Of course, since we do not have a real HERA observation of the power spectrum, we must pick a fiducial set of parameter values for our mock observation. In this paper, we adopt fiducial values $\zeta_0 = 25$, $M_{\textrm{turn}, 0} = 5\times 10^8 M_{\odot}$, $R_{\textrm{mfp},0} = 30\,\textrm{Mpc}$, and $\beta_0 = 0.936$ unless otherwise indicated. 

\section{Results}
\label{sec:MCMCresults}
%

In this section we present the results of our MCMCs and discuss their implications. We perform our computations for four different scenarios:
\begin{enumerate}
\item A HERA measurement of the 21cm power spectrum over an extended period of the EoR history, from $z = 6$ to $z = 10$.
\item Radio frequency interference (RFI) makes measurements impossible in certain frequency bands. Based on preliminary observations using HERA, we assume that a measurement can be made in a relatively clean way from $z = 7.5$ to $z = 8.5$ (using the discrete bands described in Section \ref{sec:MCMCsetup}.
\item A similar RFI-free window from $z = 9.5$ to $z = 11.5$.
\item A measurement of a $z = 8.0$ power spectrum at three separate bins centred on $k = 0.2\,\textrm{Mpc}^{-1}$, $k = 0.25\,\textrm{Mpc}^{-1}$, and $k = 0.3\,\textrm{Mpc}^{-1}$ of widths $\Delta k = $  $0.05$ Mpc$^{-1}$. These were selected by numerical experimentation to determine a minimal set of measurements required to constrain $\beta$.
\item The same measurement as scenario (ii), but assuming outside-in reionization as our fiducial model. We use the same fiducial parameters for $\zeta_0$ , $M_{\textrm{turn}, 0}$ and $R_{\textrm{mfp},0}$ as described in Section \ref{sec:MCMCsetup}, but pair it with new fiducial correlation parameter of the opposite sign, i.e., $\beta_0 = -0.936$.
\end{enumerate}
%

\subsection{Scenario (i): Measurement of $\Delta^2_{21}$ Over Redshifts $z = 6$ to $z = 10$}

\begin{figure}
  \centering
  \includegraphics[width=.49\textwidth]{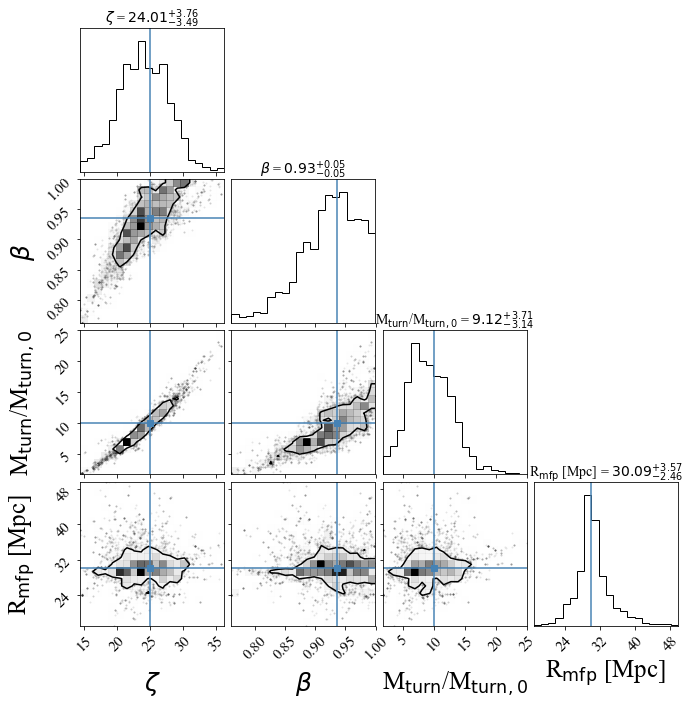}
  \caption{Posterior distributions for Scenario (i) of Section \ref{sec:MCMCresults}, covering the redshift range of our forecasts, $6 \le z \le 10$. The $68 \%$ credibility contours are entirely contained within the inside-out region of parameter space, suggesting that with such a measurement we can firmly rule out outside-in or uncorrelated models.}
  \label{fig:MCMC_z6-10}
\end{figure}

In Figure \ref{fig:MCMC_z6-10} we show our forecasts for a full HERA measurement in the range $6 < z < 10$. Immediately clear are degeneracies between certain parameters in our model. For example, since both $\zeta$ and $M_{\rm turn}$ control the timing of reionization as well as the size of the ionized bubbles at a fixed redshift, we find a considerable degeneracy between these parameters. This result is consistent with previous parameter studies such as \citet{AdrianMCMC}, \citet{GreigMesinger21cmmc2015}, and \citet{LiuParsonsForecast2016}.

Our new correlation parameter $\beta$ also exhibits slight degeneracies with $M_{\rm turn}$ and $\zeta$. The origin of this degeneracy can be deduced by studying how $\beta$ and $\zeta$ affect $\Delta^2_{21}$ at $k$ bins and redshifts where HERA is the most sensitive to changes in the model. Most of the information comes from $z \sim 8$ and $0.2\,\textrm{Mpc}^{-1} < k < 0.5\,\textrm{Mpc}^{-1}$. This redshift is slightly lower than the redshift at which the power spectrum peaks because the error bars on the power spectrum are lower at lower redshifts. The $k$ range is a balance between where the power spectrum is the most sensitive to changes in $\beta$ (see Figure \ref{fig:vert_pspec}), foreground contamination at low $k$, and instrumental noise limitations at high $k$. In this context, consider the left panel of Figure \ref{fig:z_evolution}. One sees that decreasing $\beta$ increases the amplitude of $\Delta^2_{21}$. Since $z=8$ comes after the midpoint of reionization in our fiducial model, the power spectrum is declining as one moves to lower redshifts. A decrease in $\beta$ can therefore be mimicked by a delay in reionization that shifts all the curves in Figure \ref{fig:z_evolution} to the left, bringing the power up to closer to its peak value. This can be accomplished by a decrease in $\zeta$, and indeed, one sees a positive degeneracy between $\beta$ and $\zeta$ in Figure \ref{fig:MCMC_z6-10}.

Our timing argument does not translate directly to the degeneracy between $\beta$ and $M_{\rm turn}$. Indeed, if one only considered the fact that lowering $M_{\rm turn}$ results in earlier reionzation, one would predict an opposite trend to what is seen in Figure \ref{fig:MCMC_z6-10}. However, whereas $\zeta$ mainly affects the timing of reionization, $M_{\rm turn}$ also somewhat affects the shape of the power spectrum \citep{AdrianMCMC}. Moreover, the aforementioned degeneracy between $M_{\rm turn}$ and $\zeta$ works in the opposite direction as the timing argument. It is therefore not obvious \emph{a priori} how $\beta$ and $M_{\rm turn}$ should be correlated, and Figure \ref{fig:MCMC_z6-10} reveals that the net effect is a positive degeneracy.

In any case, one sees that despite the degeneracies between $\beta$ and the other parameters, it is still possible to obtain strong constraints on the former. Examining the marginalized distributions, for instance, one sees that the $68\%$ credibility regions (CR) strongly rule out both outside-in and uncorrelated scenarios. We find that the $95\%$ CR is also entirely contained within the inside-out portion of the posterior. This allows us to rule out other reionization scenarios at $95\%$ credibility. 

\subsection{Scenarios (ii) to (iv): Measurements at Specific Redshifts and Scales}

\begin{figure}
  \centering
  \includegraphics[width=.49\textwidth]{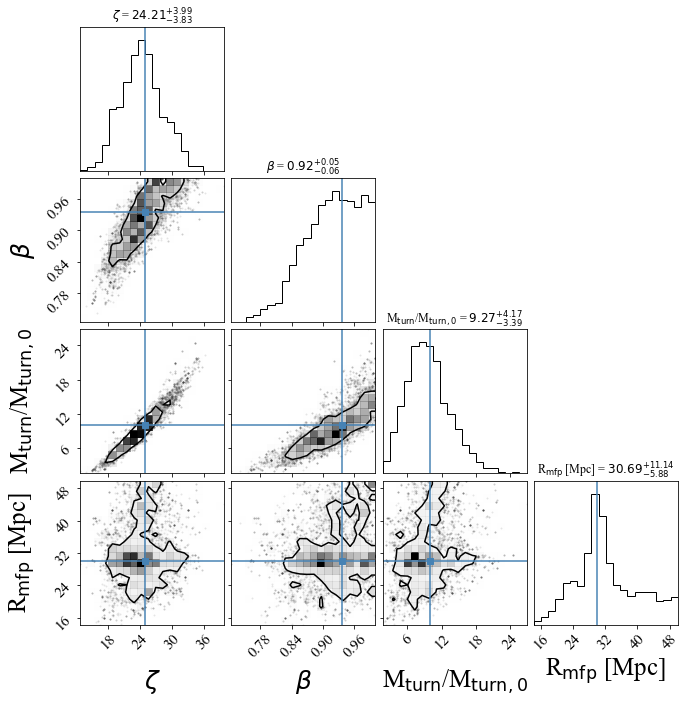}
  \caption{Same as Figure \ref{fig:MCMC_z6-10}, but for a restricted redshift range of $7.5 \le z \le 8.5$.  Measurement over these redshifts can strongly rule out correlations that are inconsistent with inside-out models. The $68 \%$ credibility contours clearly distinguish inside-out and outside in regions of parameter space. }
  \label{fig:MCMC_z7-8}
\end{figure}

\begin{figure}

  \includegraphics[width=0.49\textwidth]{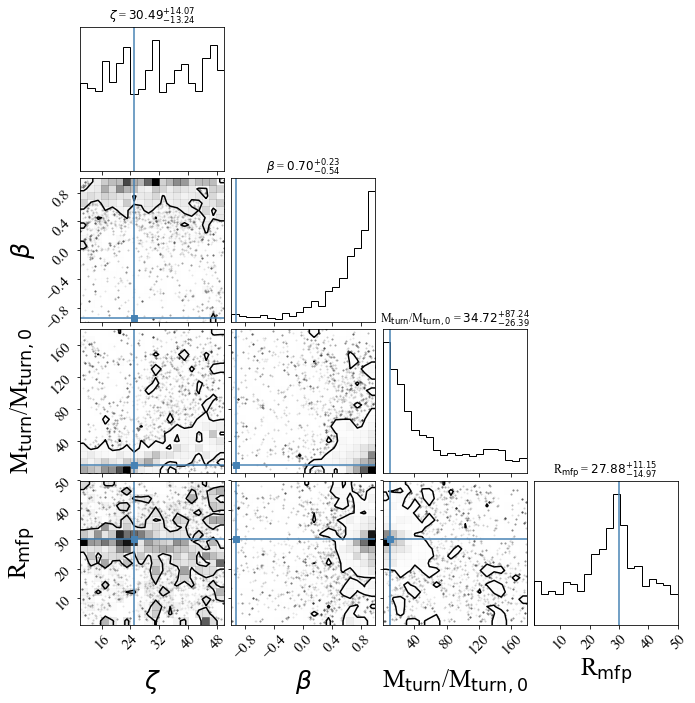}
  \caption{Same as Figure \ref{fig:MCMC_z6-10}, but for a restricted redshift range of $9.5 \le z \le 11.5$. Although many parameters are poorly constrained, an uncorrelated scenario of $\beta = 0$ can still be easily ruled out.}
\label{fig:MCMC_z9-11}
\end{figure}

Early HERA measurements will likely come at particular redshifts and $k$ bins, and in Figures \ref{fig:MCMC_z7-8} and \ref{fig:MCMC_z9-11}, we show our forecasts for the clean $7.5 \le z \le 8.5$ and $9.5 \le z \le 11.5$ spectral windows. What we find is that measurements of the power spectrum at specific key redshifts can be enough to constrain the sign of $\beta$.


\begin{figure}

  \includegraphics[width=0.49\textwidth]{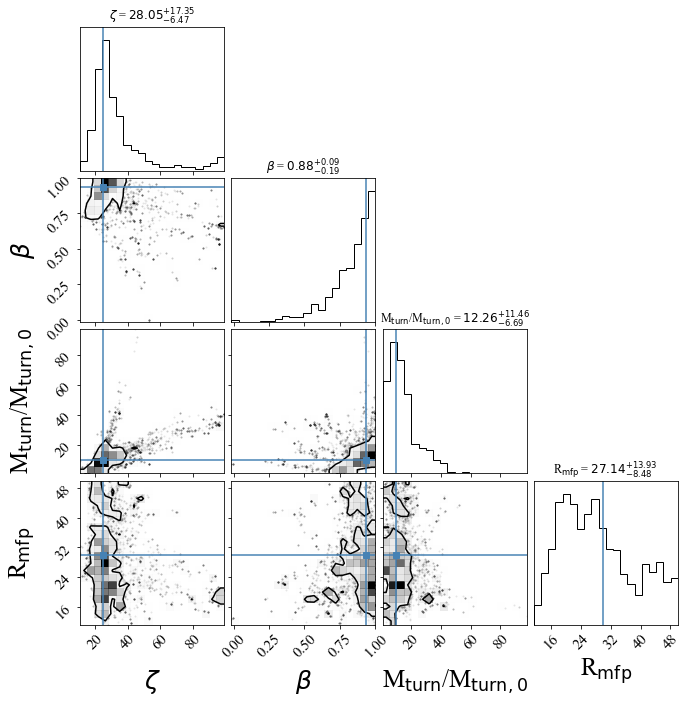}
  \caption{Same as Figure \ref{fig:MCMC_z6-10}, but using only three $k$ bins centred on $k = 0.2\,\textrm{Mpc}^{-1}$, $k = 0.25\,\textrm{Mpc}^{-1}$, and $k = 0.3\,\textrm{Mpc}^{-1}$ at $z = 8.0$. Because these bins contain the most information on $\beta$, using them alone is sufficient for excellent constraints.}
\label{fig:MCMC_z8}
\end{figure}

Consider first the $7.5 \le z \le 8.5$ window (Figure \ref{fig:MCMC_z7-8}). The conclusions drawn for this redshift window are comparable with the results over the expanded history of the EoR (i.e., the previous scenario with measurements from $z = 6$ to $10$). This is unsurprising since the bulk of the information of $\beta$ comes from precisely these redshifts: recall from Figure \ref{fig:z_evolution} that this range coincides to redshifts where the power spectrum peaks, as well as where the differences between different $\beta$ are the greatest. Indeed, to take this to an extreme, Figure \ref{fig:MCMC_z8} shows the constraints from measuring just three $k$ bins (centred on $0.2\,\textrm{Mpc}^{-1}$, $0.25\,\textrm{Mpc}^{-1}$, and $0.3\,\textrm{Mpc}^{-1}$) at redshift $z = 8.0$. These are the $k$ and $z$ bins that contain the most information about $\beta$. The results are again similar. Of course, the constraints are slightly worse than one obtains over the full redshift range. However, the differences are quantitative rather than qualitative, and one can easily distinguish between inside-out and outside-in models of reionization as the $\beta$ contours are entirely contained within the inside-out region of parameter space. We find that our ability to distinguish between these models is also valid at $95\%$ CR. 

With the $9.5 \le z \le 11.5$ window,\footnote{For this scenario only, we exclude $k > 0.75\,\textrm{Mpc}^{-1}$ for computational simplicity. There is little sensitivity to the power spectrum here anyway, and thus there is a negligible change to our results.} we probing the early stages of reionization, where the process has yet to make a significant imprint on the temperature field. The power spectrum is therefore less sensitive to $\beta$.
Correspondingly, these redshifts are not as effective at numerically constraining the value of $\beta$; however, we can still confidently determine the sign of $\beta$. Outside-in morphologies, for example, are inconsistent with our mock observation of a fiducial inside-out model. 
Importantly, we note that this is true even though the limited redshift range is unable to provide strong constraints on other parameters, although it is important to acknowledge that for this scenario the constraint on $\zeta$ is mostly driven by our prior.

In principle, the value of $\beta$ need not be constant in time, and could vary throughout the EoR history. In this scenario $\beta$ would obtain a redshift dependence. Given this reality, one can then confine measurements of $\beta$ to epochs where the correlations between $\delta$ and $x_{\rm HII}$ remain constant. For example, our forecast for $\beta$ between $7.5 < z < 8.5$ is entirely contained within the ``Ionized Fibre" stage of reionization in \cite{Cheng2019}. Since forecast scenarios (ii) through (iv) are small enough redshift windows, they each fall within a particular ``stage" of reionization morphology. Making the correspondence between each of these scenarios and a particular stage of reionization morphology becomes more difficult for scenario (i) which takes place over $6 < z < 10$.

\subsection{Scenario (v): Fiducial Outside-in Reionization}
 \label{sec:outsidein}
\begin{figure}

  \includegraphics[width=0.49\textwidth]{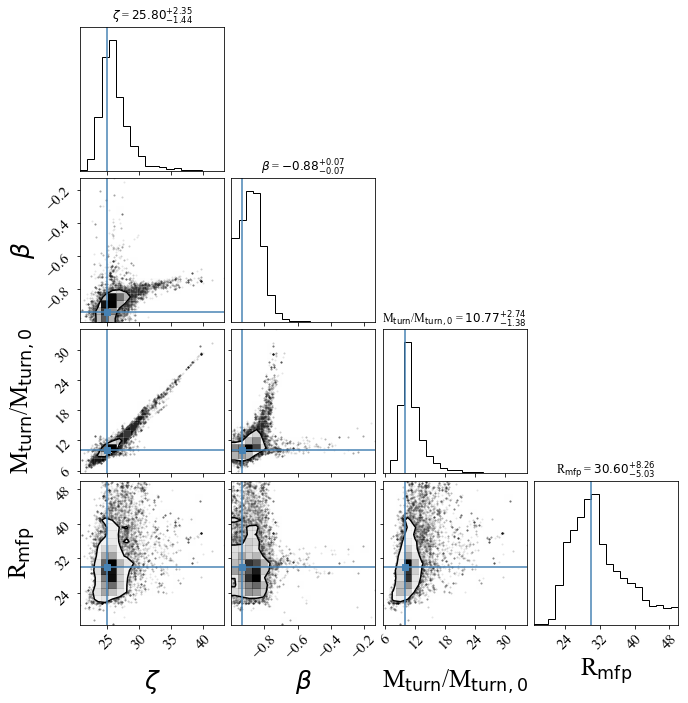}
  \caption{Same as Figure \ref{fig:MCMC_z7-8}, but using an outside-in model with $\beta = -0.936$ as our fiducial reionization morphology. Displayed are the $68 \%$ CR contours. The  $68 \%$ CR are entirely contained within $\beta < 0 $ suggesting that if reionization proceeds as outside-in, measurement of $\Delta^2_{\rm 21}$ in this scenario can rule out uncorrelated and inside-out reionization with $68\%$ credibility.}
\label{fig:MCMC_outsidein}
\end{figure}

 In figure \ref{fig:MCMC_outsidein}, we show the MCMC posterior for measurement of $\Delta^2_{\rm 21}$ with parameters identical to scenario (ii), but using a fiducial outside-in morphology ($\beta_0 = - 0.936$). In an outside-in scenario, $\Delta^2_{\rm 21}$ varies slowly as a function of $\beta$ in the regime $\beta \simeq -1$ (see Figure \ref{fig:vert_pspec}). As a result, it is difficult to numerically distinguish the particular value of $\beta$ for extreme outside-in models. However this does not prevent us from being able to distinguish outside-in models from uncorrelated or inside-out reionization. The 1D posterior for $\beta$ is non-Gaussian around $\beta = -1$ and the posterior drops dramatically around $\beta \simeq -0.7$, which is where $\Delta^2_{\rm 21}$ becomes very sensitive to changes in correlation.  This allows us to properly distinguish between the signs of $\beta$. The $68\%$ CR contours of such a measurement are entirely contained within $\beta < 0$.  Therefore for this fiducial reionization scenario, measurements of $\Delta^2_{\rm 21}$ across the midpoint of reionization can fully distinguish between signs of $\beta$. This conclusion is also valid at a $95\%$ CR level. That is, the $95\%$ credibility region is entirely contained within the outside-in regime of the posterior.
 
\subsection{Early Limits}


 In closing, we note an interesting trend that will likely manifest as upper limits from early data slowly come down. From Figure \ref{fig:z_evolution}, note that the brightest power spectra come from outside-in scenarios. Thus, early upper limits will first rule out (or detect!) these morphologies. As we alluded to in Section \ref{sec:outsidein}, however, the curves for $\beta < 0$ (outside-in) are typically closer together than those for $\beta > 0$. Therefore, although the outside-in scenarios are the easiest to rule out or detect, a generic quantification of the density-ionization correlation is easier (i.e., the error bars on $\beta$ will be smaller) if our Universe ends up being one where reionization proceeded in an inside-out fashion. 

\begin{figure}
  \includegraphics[width=0.4\textwidth]{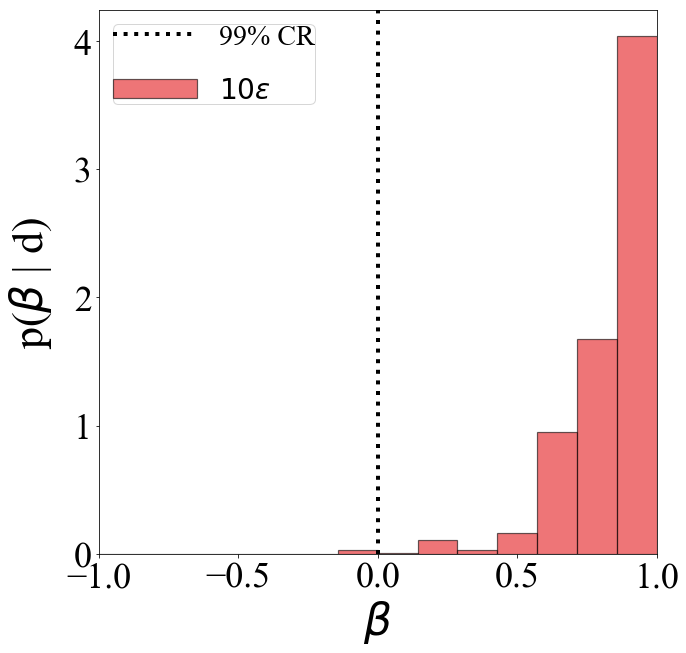}
  \caption{Marginalized posterior distributions on $\beta$ obtained through the measurement of $\Delta^2_{21}$ in Scenario (ii) of Section \ref{sec:MCMCresults}, but assuming 10 times larger error bars than the design HERA sensitivities to mimic potential upper limits. As the upper limits come down to 10 times the fiducial errors on $\Delta^2_{21}$ from Section \ref{eq:sensitivities}, we will be able to rule out $\beta < 0$ with 99\% credibility in our fiducial model. The dotted vertical lines demarcate $\beta = 0$ to guide the eye.} 
\label{fig:1Dcontours}
\end{figure}

We expect that as upper limits descend to roughly $10$ times the design sensitivity of HERA (i.e., error bars approach $\sim 10 \varepsilon$ on a measurement of $\Delta^2_{21}$), we will be able to rule out uncorrelated reionization ($\beta=0$) at $\sim 99\%$ credibility, assuming the same $z$ and $k$ bins as Scenario (ii) in Section \ref{sec:MCMCresults}. This is illustrated in Figure \ref{fig:1Dcontours} which shows the posterior on $\beta$ from upper limits where the power spectrum is constrained with errors on the order of $\varepsilon \sim\!\!20\,\textrm{mK}^2$ (corresponding to $10$ times the HERA design errors), for bands described in Section \ref{sec:MCMCsetup}.


\section{Conclusion}
\label{sec:Conclusion}
The first measurements of the 21cm power spectrum are upcoming and a key prediction of many EoR models is the correlation between ionization and density fields. The way these two fields correlate is an indicator of the morphology of reionization. A number of discrete cases of correlations have been studied in previous works pertaining to inside-out or outside-in models. We build upon these works by proposing a continuous parameter $\beta$ that encompasses all possible correlation scenarios to provide an alternative method for model selection.

A key effect of changing $\beta$ in 21cm temperature maps is to alter the contrast between ionized bubbles and neutral regions. The temperature contrast between these regions is larger for values of $\beta < 0$, corresponding to outside-in models of reionization. Inside-out models, with $\beta > 0 $, have decreased contrast. Similarly, we find that outside-in models of reionization produce an increase in the amplitude in the 21cm power spectrum compared to their inside-out counterparts. The distinction between these models is greatest at the midpoint of reionization. 

Upcoming limits on the 21cm power spectrum will allow us to place constraints on $\beta$, potentially ruling out or favouring various models. As a test case, we carry out a numerical forecast using the HERA experiment, which has the sensitivity to place $> 20\sigma$ limits on $\Delta^2_{21}$. We perform an MCMC over the entire history of reionization and find that the fiducial inside-out model used produces a unique imprint on the 21cm power spectrum. Measurement of $\Delta^2_{21}$ over this redshift range can fully distinguish between the two classes of correlations. Because measurements of $\Delta^2_{21}$ over a wide range of $z$ and $k$ bins are unlikely in early observations, we select a smaller range of redshifts to measure $\Delta^2_{21}$. Motivated by the high sensitivity of $\Delta^2_{21}$ on $\beta$ at the midpoint of reionization, we identify redshift $z \sim 8$ as the most effective regime to distinguish between types of correlations. Measurements in just a few select $k$ bins at these redshifts are
sufficient for learning about the broad morphology of reionization. Along the way, such measurements will test a key measure of reionization---the degree of correlation between the density and ionization fields---laying the groundwork for increasingly detailed $21\,\textrm{cm}$ constraints that will considerably enhance our knowledge of the EoR.

\section*{Acknowledgements}

The authors are delighted to acknowledge helpful discussions with Phil Bull, Ad\'{e}lie Gorce, Jordan Mirocha, and Steven Murray, as well as Steve Furlanetto for posing the original question that inspired this work.  We acknowledge support from the New Frontiers in Research Fund Exploration grant program, a Natural Sciences and Engineering Research Council of Canada (NSERC) Discovery Grant and a Discovery Launch Supplement, the Sloan Research Fellowship, the William Dawson Scholarship at McGill, as well as the Canadian Institute for Advanced Research (CIFAR) Azrieli Global Scholars program. This research was enabled in part by support provided by Calcul Quebec (\url{www.calculquebec.ca}), WestGrid (\url{www.westgrid.ca}) and Compute Canada (\url{www.computecanada.ca}). The authors would like to thank the anonymous referee for the insightful comments which have improved this paper.



\section*{Data Availability}
The software code underlying this article will be shared on reasonable request to the corresponding author.

\bibliographystyle{mnras}
\bibliography{example} 




\appendix


\bsp	
\label{lastpage}
\end{document}